\documentclass[%
reprint,
superscriptaddress,
 amsmath,amssymb,
 aps,
prl,
]{revtex4-2}
 
\usepackage{float}
\usepackage{xcolor}
\usepackage{graphicx}
\usepackage{dcolumn}
\usepackage{bm}

\begin{document}

\title{An elasto-plastic approach based on microscopic insights for the steady state and transient dynamics of sheared disordered solids}
\author{Chen Liu}
\affiliation{
Laboratoire de Physique de l'Ecole Normale Supérieure, Paris, France
}%
\author{Suman Dutta}
\affiliation{
The Institute of Mathematical Sciences, Taramani, Chennai 600113, India
}%
\author{Pinaki Chaudhuri} 
\affiliation{
The Institute of Mathematical Sciences, Taramani, Chennai 600113, India
}%
\author{Kirsten Martens}
\affiliation{
Univ. Grenoble Alpes, CNRS, LIPhy, 38000 Grenoble, France
}%

\date{\today}

\begin{abstract}
We develop a framework to study the mechanical response of athermal amorphous solids via a coupling of mesoscale and microscopic models. Using measurements of coarse grained quantities from simulations of dense disordered particulate systems, we present a coherent elasto-plastic model approach for deformation and flow of yield stress materials. For a given set of parameters, this model allows to match consistently transient and steady state features of driven disordered systems with diverse preparation histories under both applied shear-rate and creep protocols. 
\end{abstract}

\maketitle

\noindent
{\it Introduction.} Amorphous materials under deformation exhibit a wide spectrum of non-trivial phenomena, that not only elicit fundamental questions, but also bring about challenges {in} engineering \cite{rodney2011modeling, bonn2017yield, nicolas2018deformation, cipelletti2020microscopic}). 
One of the major goals in this context, is to develop a unique theoretical framework {which describes} transient phenomena like stress overshoots prior to yielding (e.g.~in metallic glasses
\cite{kawamura1999stress, lu2003deformation, maass2012shear} 
and soft materials 
\cite{amann2013overshoots, divoux2011stress, dzuy1983yield, rogers2010time, zausch2008equilibrium, sentjabrskaja2014transient, koumakis2012yielding, koumakis2012direct}), delayed failure in creep experiments \cite{chaudhuri2013onset, sentjabrskaja2015creep, cipelletti2020microscopic}, along with steady state properties, e.g. strongly non-linear flow curves \cite{agoritsas2017non}.

It is well established that deformation of disordered materials is realised through successive dissipative events in the form of localised shear transformations \cite{argon1979plastic}, resulting in long range elastic stress variations in the surroundings \cite{eshelby1957proc}, potentially leading to {cascades of plastic deformations} 
\cite{baret2002extremal, martens2011connecting}. 
Following this very generic picture, it has been proposed that
brittle amorphous materials, such as metallic glasses, and soft disordered matter, such as emulsions or colloidal suspensions, can be described by similar mesoscopic modeling approaches \cite{nicolas2018deformation}. To reveal the underlying physics and unify the understanding of the various phenomena in yielding and flow of amorphous systems, it is thus tempting to derive models on the mesoscopic scale, using coarse grained quantities like a local tensorial stress, strain and corresponding elastic moduli \cite{tsamados2009local}.

In this spirit, several coarse grained scenarios have been developed such as the soft glassy rheology model \cite{sollich1997rheology, moorcroft2011age}, fluidity models \cite{picard2002simple, bocquet2009kinetic}, the shear transformation zone theory \cite{falk1998dynamics}, the mode-coupling theory \cite{fuchs2002theory} and a large number of elasto-plastic descriptions \cite{hebraud1998mode, baret2002extremal, rodney2011modeling,merabia2016thermally,nicolas2018deformation}. Although some of these models could be successfully fitted to simulations and experiments \cite{manning2007strain, goyon2008spatial,  zausch2008equilibrium, mansard2013molecular,nicolas2013spatial}, the ingredients of these mesoscopic models remain in most cases phenomenological and the direct link to underlying microscopic dynamics remains unresolved.

Attempts to infer directly coarse grained parameters from particle based simulations for effective large scale descriptions are indeed rather scarce. For example, in a recent work, an effective temperature is inferred on various coarse graining scales to rationalise the formation of transient shear bands \cite{hinkle2017coarse}. On the other hand, within the framework of elasto-plastic models that involve an elastic kernel in form of an Eshelby response, some links to microscopic dynamics have been explored \cite{puosi2014time, agoritsas2015relevance, nicolas2014spatiotemporal, albaret2016mapping, boioli2017shear}, including estimation of some input parameters, such as yield stress distributions \cite{puosi2015probing, patinet2016connecting, barbot2018local, barbot2020rejuvenation}. However, all these studies have usually concentrated on some specific aspects of the yielding process and there is yet no unique framework for the description of the various common protocols and the resulting phenomena.

In this letter, {we demonstrate a successful coupling of}
{microscopic} simulations and a meso-scale elasto-plastic model, {to} systematically match various steady state and transient properties of the shear response a soft disordered solid,
for different loading protocols and a large range of driving parameters, 
 using initial states with diverse preparation histories. Thereby, we validate further 
the use of simple Eshelby based elasto-plastic descriptions for the 
the deformation and flow of amorphous materials. Our results show that this procedure allows to extract the relevant physical ingredients on a coarse grained level to describe a multitude of emergent macroscopic phenomena.

\noindent
{\it Microscopic model.}  We first describe the particle based molecular dynamics (MD) simulations onto which we map and validate our coarse grained elasto-plastic model, as discussed later. We consider a well-studied two-dimensional binary Lennard-Jones (LJ) glass-former, with particles having same mass ($m$) but having a size ratio of $\frac{1+\sqrt{5}}{4}$ \cite{lanccon1988two, lanccon1986thermodynamical, falk1998dynamics,si}, using a system of size $N=102400$ in a box of length $L=316.174$ with periodic boundary conditions. We study the shear response of athermal amorphous states which are inherent structures (IS) \cite{sastry1998signatures} of this model residing at energy levels $E_1=-2.3254, E_2=-2.3439, E_3=-2.3771$, respectively obtained via quenches from the super-cooled and glassy regime; see SI \cite{si} for further details. Here, the units of length, energy and time are respectively $\sigma_{\rm LS}$,  $\epsilon_{\rm LS}$ and $\sqrt{{m\sigma_{\textrm{LS}}^{2}} /
\epsilon_{\textrm{LS}}}$; all observables are measured in these LJ units \cite{si}. The athermal shear response of these IS states is probed by imposing two different protocols, viz.~constant shear-rate
and constant shear stress. For the applied shear-rate, the simulation box is deformed in the $xy$ plane
at a rate determined by the imposed shear-rate, and in parallel the particles' equations of motion are numerically
integrated.  Microscopic dissipation is controlled via a term in the equation of motion of each particle,  viz.~$-\zeta ({{\bf r}_{ij}} ~. ~{\bf v}_{ij})\hat{{\bf r}}_{ij}$, where $r_{ij}$ is the inter-particle distance, $v_{ij}$ is the relative velocity and $\zeta$ is the microscopic dissipation coefficient \cite{si}.
The response to an applied shear stress is {implemented} via a feedback method \cite{vezirov2015manipulating, cabriolu2019precursors}, wherein, apart from the integration of the equations of motion of the constituent particles, which includes the dissipation term discussed above, we also integrate simultaneously the equation for the {time evolution} of the macroscopic shear rate ($\dot{\gamma}$): $d\dot{\gamma}(t)/{dt}=B[\sigma_0-\sigma_{xy}(t)]$, where $\sigma_0$ is the imposed target stress, $\sigma_{xy}$ is stress developing inside the system due to the external drive and $B$ is a damping coefficient. For our study, we choose $B=1$, $\zeta=1$ \cite{cabriolu2019precursors}. In all cases, integration of MD equations are done using a time-step of $0.005$, via LAMMPS \cite{plimpton1993fast}.

\noindent
{\it Elasto-plastic model.}
To coarse grain the dynamics we use a lattice based scalar elasto-plastic model. Each site of the lattice represents a typical particle cluster that undergoes plastic events (or shear transformations) quantified by a local plastic strain $\gamma^{pl}_i$.
 Assuming linear elasticity, a site $i$ sustains a local stress 
$\sigma_i=\mu\gamma^{el}_i$ with $\mu$ the shear modulus.
Due to the response to local plastic events, the local stress $\sigma_i$ fluctuates around its spatial average, the externally applied stress  
$\Sigma^{\text{EXT}}\equiv\sum_i\sigma_i/N$. The coupling between the stress fluctuations and the plastic strain field is formulated by
\begin{equation}\label{eq:sigma}
    \sigma_i =\Sigma^{\text{EXT}} + \mu \sum_j G_{ij}\gamma^{pl}_j=\mu\gamma^{el}_i
\end{equation}
where $G_{ij}\propto \cos(4\theta)/r^2$ is the Eshelby kernel satisfying $\sum_i G_{ij}=0$.  The relaxation dynamics of a plastically deforming site $i$ is given as 
$\partial_t \gamma^{pl}_i = n_i \sigma_i/(\mu\tau)$
where we fix $\tau=1$ as the intrinsic relaxation time scale \cite{puosi2015probing}.
 The stochastic dynamics of the local state variable $n_i$ 
reads as follows.
A site becomes plastic from an elastic state 
$n_i=0\rightarrow n_i=1$ at a rate $1/\tau^{pl}$ only if it exceeds a local threshold, i.e. $\sigma_i>\sigma_i^{\text{th}}$. Once a site is in a plastic state, the local elasticity is recovered at a rate $1/\tau^{res}$ and a new local threshold is randomly generated from a prescribed 
threshold distribution $P_d(\sigma^{\text{th}})$. The equations of motion have been integrated using a time step $\Delta t = 10^{-2}$.

The above rules together with a protocol for $\Sigma^{\text{EXT}}(t)$ 
and an initial condition $\{\gamma^{pl}_i(t=0),\sigma^{\text{th}}_i(t=0)\}_i$ allow for the implementation of a creep protocol, by fixing $\Sigma^{\text{EXT}}$ to the desired value of the imposed stress \cite{liu2018creep}.  And a fixed shear rate protocol is realized by controlling $\Sigma^{\text{EXT}}(t)$ through a feedback loop 
\begin{equation}
    \frac{d}{dt}\Sigma^{\text{EXT}}(t)=\mu\left( \dot{\gamma} - \frac{1}{N}\sum_i \dot{\gamma}^{pl}_i(t;\Sigma^{\text{EXT}})  \right)
\end{equation}
in such a way that the macroscopic shear rate $\dot{\gamma}=\frac{1}{N}\sum_i (\dot{\gamma}_i^{pl}+\dot{\gamma}^{el}_i )$ remains at the desired constant value. 

The mesoscopic model accounts for the rheological behavior of amorphous materials as a result of the interplay of three parts : (i) the elasto-plastic dynamic rules, (ii) the loading condition $\Sigma^{\text{EXT}}(t)$, and (iii) the initial condition.
In the microscopic model this initial condition is determined by the annealing protocol.
Our main assumption is that we capture all relevant features of the initial condition, knowing the local threshold map $\{\sigma^{th}_i\}$, the stress map $\{\sigma_i\}$  and the shear modulus $\mu$ in the initial state \cite{si}.
Thus, as illustrated later, any variation in the initial state only need changes in these inputs,  
sufficient to describe its effect on various rheological behaviors, leaving all else parameters unchanged.

To provide input for the initial condition to the mesoscopic model, we divide our MD simulation box into a square grid and measure the coarse-grained stress and threshold fields using a frozen matrix method \cite{sollich2011cecam, patinet2016connecting, barbot2018local, si}. This procedure introduces a coarse-graining {scale, which} represents the typical number of particles involved in an Eshelby inclusion. The input shear modulus is measured as the slope in the stress-strain curve upon small deformations. If our approach is correct, there should exist a unique set of model parameters such that the elasto-plastic model captures various simulation observations by only plugging the corresponding loading and initial conditions assessed from the microscopic model.

We set the typical time of stress relaxation $\tau=1$ \cite{puosi2015probing}, which coincides with the time unit of the microscopic model. For the coarse-graining length, we use approximately $9.88 \sigma_{\rm LS}$, similar to what was used recently in computing local yield stresses \cite{patinet2019origin}.
Further we tune the other parameters, namely $\tau^{pl}$, $\tau^{res}$ and adjust the form of $P_d(\sigma^{th})$ to match the meso model to our microscopic simulations. We choose $P_d(\sigma^{th})$ to be a Weibull distribution \cite{patinet2016connecting,patinet2019origin,si} 
\begin{equation}
w_d(\sigma^{th};k_d,\sigma^{th}_d)=\frac{k_d}{\sigma^{th}_d}\left(\frac{\sigma^{th}}{\sigma^{th}_d}\right)^{k_d-1}\exp\left[-\left(\frac{\sigma^{th}}{\sigma^{th}_d}\right)^{k_d}\right]
\end{equation}
where $k_d$ controls its shape and $\sigma^{th}_d$ specifies the typical value of thresholds. 

To find the best set of parameters, we first compare observables in the stationary state, which is independent of
initial conditions. Then, transient dynamics is compared, starting from the imposition of shear,
for two different shear protocols, viz.~constant shear-rate and constant shear stress. Combining empirical tuning and some quantitative analysis of the effects of the parameters\cite{si}, we  find a setting of parameters that allows to reproduce systematically the MD results. Via this, we fix the parameters of the meso scale description to $\tau^{pl}=1$, $\tau^{res}=2.0$, $k_d=1.5$, $\sigma^{\text{th}}_d=0.57$, which allows for very good quantitative match between the meso model and the microscopic simulations, for several observables, discussed in details below.

\begin{figure}[]
\begin{center}
\includegraphics[width =\columnwidth]{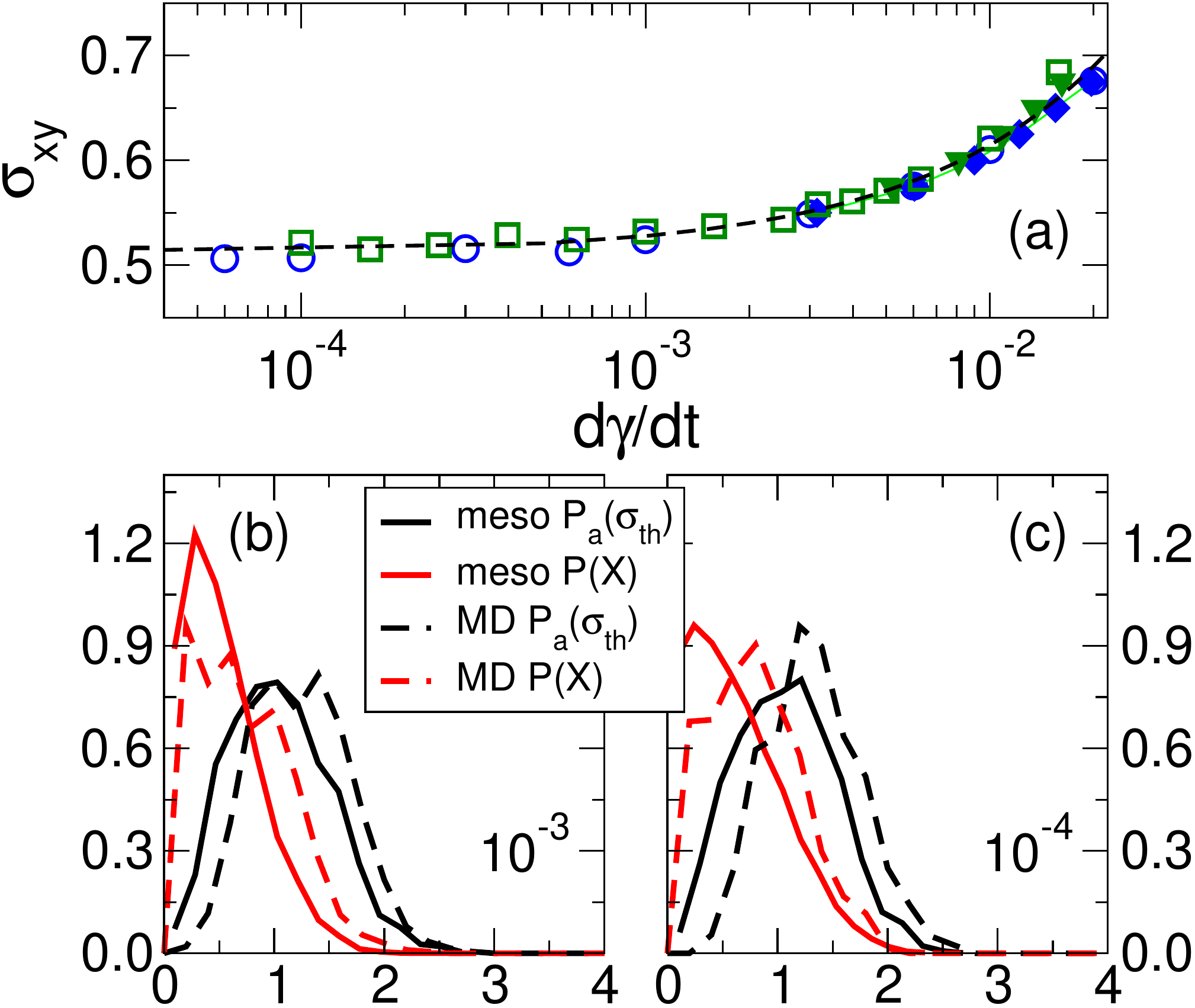}
\caption{(a) Comparison of the flow curve (shear stress $\sigma_{xy}$ vs shear-rate $d\gamma/dt$) obtained from the mesoscale (in green) and microscopic (in blue) models using constant shear-rate (empty symbols) and constant stress (filled symbols) protocols. Dashed line is a Herschel-Bulkley fit with yield stress $\Sigma^Y=0.5107$ and $n=0.784$ (see main text). (b,c) Histograms of local yield threshold $\sigma_{th}$ and local distance to yield $X$, measured in steady state, for the microscopic  (dashed line) and the elastoplastic (filled bars) models, at imposed shear rates $\dot{\gamma}=10^{-3}$ (left panel) and $\dot{\gamma}=10^{-4}$ (right panel).} 
\label{fig:fc}
\end{center}
\end{figure}

\noindent
{\it Flow curve.} The stationary flowing state, attained either by imposing a fixed shear-rate or a fixed shear-stress, is characterized by a unique relationship between the shear stress $\sigma$ and the shear rate $\dot{\gamma}$. In Fig.\ref{fig:fc}(a), we show flow curves obtained via both protocols using data from microscopic simulation and mesoscopic model, which match consistently over two decades in strain rate. For imposed shear-stress, 
the stress barrier for attaining steady state depends on the initial condition of the system and is potentially larger than the dynamic yield stress $\Sigma^Y$ {(stress value in the limit $\dot{\gamma}\rightarrow{0}$)} \cite{rodney2011modeling,ozawa2018random}, which explains why the data from the creep protocol is limited (see Fig.\ref{fig:fc}(a)).
The {obtained} flow curves can be fitted by a Herschel-Bulkley fit $(\Sigma-\Sigma^Y) \sim \dot{\gamma}^n$ with $n<1$; see Fig.\ref{fig:fc}(a). For shear rates larger than $\dot{\gamma}_X\approx 0.02$ (not shown in Fig.1(a)), the meso model assumptions no longer hold and do not reproduce correctly the steady flow \cite{si}.

\noindent
{\it Apparent distribution of thresholds $P_a(\sigma^{\text{th}})$.} A frequently studied observable characterizing the steady state flow is the apparent threshold distribution $P_a(\sigma^{th})$ as well as gap distribution $P(X)$. In the mesoscopic model, several uncorrelated snapshots of the threshold field $\sigma^{th}_i$ are taken to construct the histogram of thresholds $P_a(\sigma^{th})$ from the steadily flowing state. Similarly, by defining the gap field as $x_i=\sigma^{th}_i-\sigma_i$, one constructs the steady state gap distribution $P(X)$ for the meso model. Such measurements are also done in the microscopic simulations, by sampling configurations in the steady state under imposed shear, and then using the frozen matrix method \cite{sollich2011cecam, puosi2015probing, patinet2016connecting, barbot2018local} to obtain the
local yield thresholds.
Results gathered from the microscopic and mesoscale simulations are plotted in Fig.1(b,c) for shear rates $\dot{\gamma}=10^{-4}$ and $\dot{\gamma}=10^{-3}$. 
When comparing the distributions, we note that the distribution obtained from the microscopic dynamics is shifted to higher values. The origin may lie in the frozen matrix method, that constrains the relaxation 
and is therefore likely to over-estimate the yield threshold. Nevertheless, the observations from the mesoscale and microscopic analysis are in good qualitative agreement and consistent. The mean of the distribution increases systematically with decreasing shear-rate. 

\begin{figure}[t]
\begin{center}
\includegraphics[width =1.0\columnwidth]{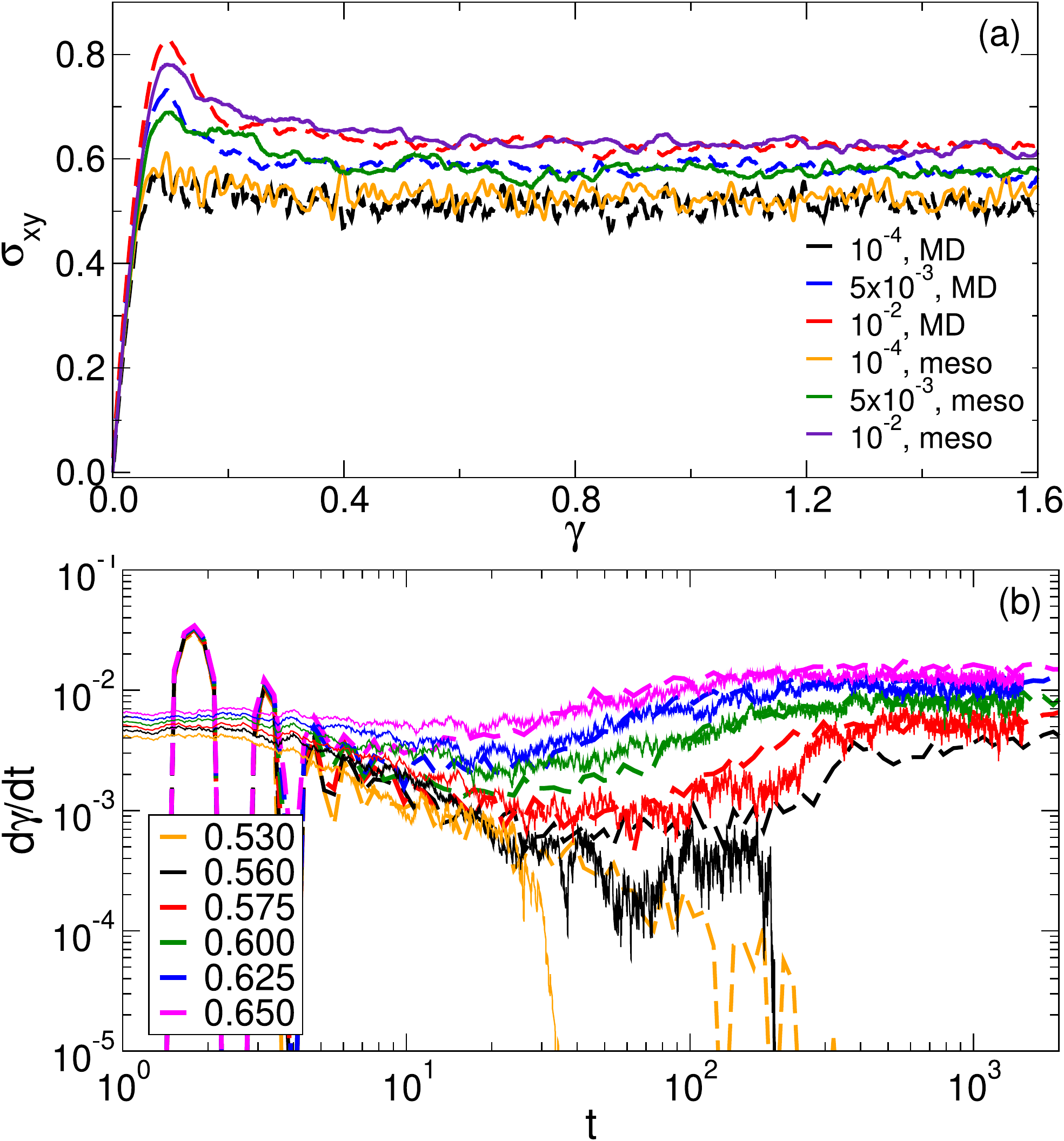}
\end{center}
\caption{(a) Comparison of the macroscopic shear stress ($\sigma_{xy}$) versus strain ($\gamma$) at three different imposed shear rates for the mesoscale and microscopic models, using a initial state having energy level $E_1$. (b) Comparison of the shear rate response to an applied step stress, for the same initial state,
over a range of magnitudes as marked, for microscopic model (in dashed lines) and mesoscale model (in filled lines).} 
\label{fig:load}
\end{figure}

\noindent
{\it Transient response.} 
To probe the response of the model systems prior to the {onset of} steady state flow, we consider the response of one of the amorphous states, viz.~the one labelled $E_1$ to different imposed shear-rates, the stress-strain curves for which are shown in Fig.\ref{fig:load}(a). 
The meso model predictions and the microscopic data compare fairly well over a broad range of shear rates. Notably, the dependence of the stress overshoot on the shear rate is well captured and the agreement of the stress level attained in the long time limit is guaranteed by the consistency of the flow curve presented in Fig.\ref{fig:fc}. 

Next, we focus on the transient regime of the same amorphous state for an imposed shear stress (see  Fig.\ref{fig:load}(b)). 
We observe typical creep responses, as
reported in previous works \cite{siebenburger2012creep, divoux2011stress, coussot2006aging, liu2018mean}. 
The striking point is that once we tune the parameters of the meso-scale model to obtain the quantitative comparisons discussed above, the agreement in the transient regime of the creep curves (Fig.\ref{fig:load}(b)) is automatically achieved, for all values of applied stress. Note that apparent differences in the short time dynamics are easily understood. The large oscillations in the early stage of the microscopic simulations origins from our feedback protocol \cite{cabriolu2019precursors}, whereas the meso-model imposes the target stress instantaneously.

\begin{figure}[t]
\begin{center}
\includegraphics[width =1.0\columnwidth]{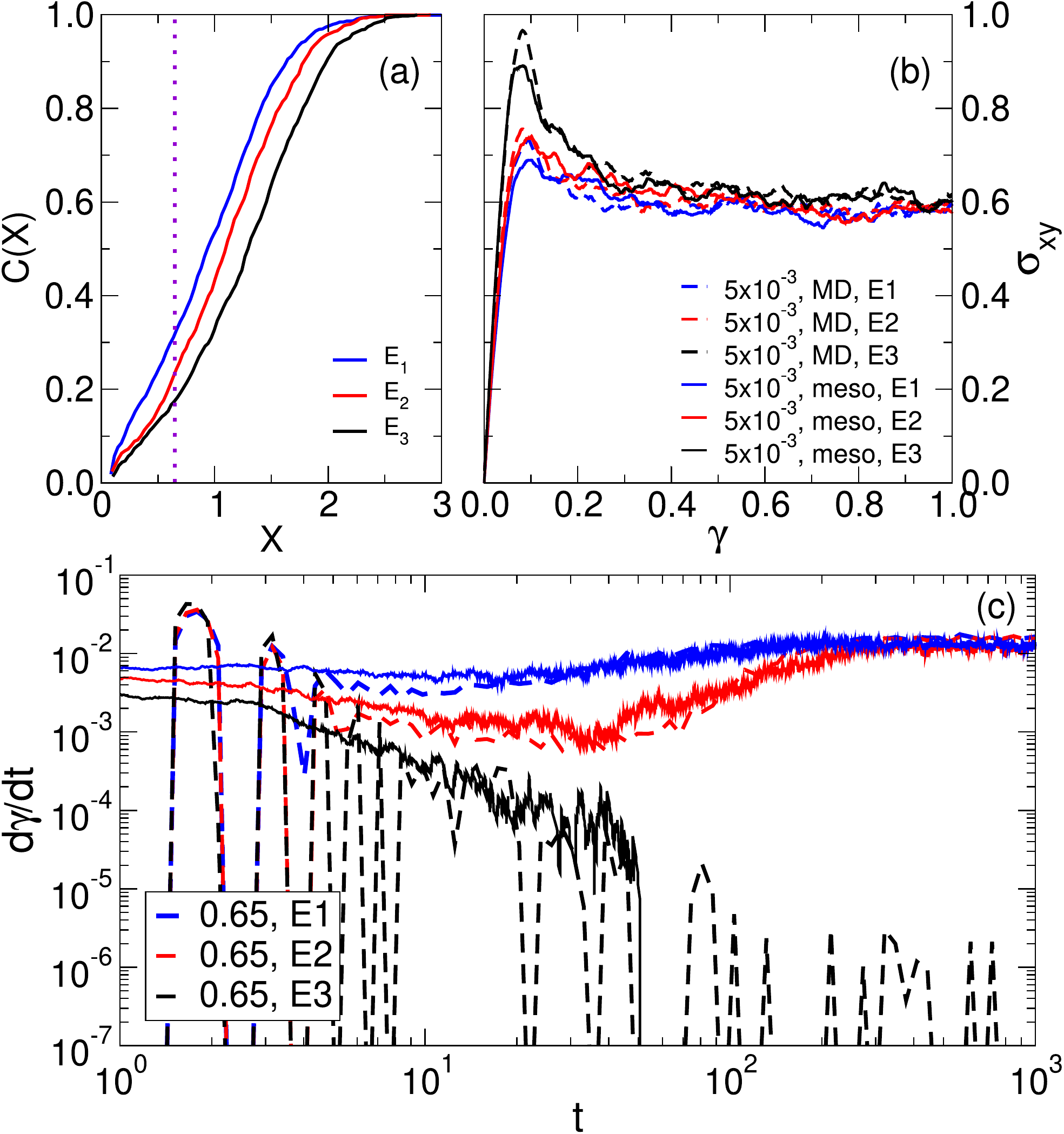}
\end{center}
\caption{(a) Cumulative distribution, $C(X)$ of local distance to yield ($X=\sigma^{\rm th}-\sigma$) for three different initial states having energies $E_1 > E_2 > E_3$. Comparative response 
of these states to (b) imposed shear-rate of $5\times{10^{-3}}$ and (c) imposed shear stress ($0.65$, marked in (a)), computed from the microscopic and mesoscale model, as labelled.} 
\label{fig:load2}
\end{figure}

For the imposed stress of $0.56$, the mismatch between the two models can be attributed to the coarse-graining procedure. Some of the microscopic fluctuations get averaged out in the elasto-plastic description, while coarse-graining, and the dynamics gets stuck in the absorbing non-flowing state more easily, unlike the microscopic model where the steady flow can still be reached. Future works should aim to develop stochastic finite element models accounting for additional noise terms for an improved description of the dynamics in the vicinity of the yielding thresholds. 

For an applied stress of $0.53$, both models predict a stuck state; the stress magnitude is smaller than the static yield threshold for this {initial state} \cite{chaudhuri2012inhomogeneous}.
 
We checked that the creep curves can also be reproduced if the mesoscale equations are initialised with random local yield thresholds, sampled from the distribution corresponding to this preparation history.
The reason is that 
the flow-onset stress is large with respect to the dynamical yield stress, implying that a large fraction of sites destabilize immediately upon the application of the external stress and potential spatial correlations in the initial state play a minute role. 

\noindent
{\it Role of initial states.} Finally, we address how the shear response of amorphous solids depends upon their preparation history. 
With decreasing energy of the IS states ($E_3 < E_2 < E_1$), macroscopic shear modulus increases \cite{si} and the local yield thresholds also increase, which is reflected in the cumulative gap distribution, $C(X)$, shifting to the higher side (see Fig.\ref{fig:load2}(a)), consistent with earlier observations \cite{ozawa2018random}. Consequently, for imposed shear-rate, we observe an increasing stress overshoot in the stress-strain response, as the energy of the initial state decreases; see Fig.\ref{fig:load2}(b).  Further, in the response to imposed stress (see Fig.\ref{fig:load2}(c)), a more prominent minimum in $\dot{\gamma}(t)$ and a sharper approach to steady state is observed, with decrease in energy of the IS state, consistent with observations in aging emulsions \cite{siebenburger2012creep}. For $E_3$, the system gets eventually stuck, indicating that the static yield threshold is higher than the imposed stress ($0.65$). Notably, in this case, a smaller fraction of sites have $X$ values lower than the imposed stress, as compared to $E_1, E_2$ (see Fig.\ref{fig:load2}(a)). Overall, the most striking observation is that the 
varied transient response, starting from these three different initial states, is well-matched for the meso-scale and microscopic models, underlining the effectiveness of the elasto-plastic modeling scheme.

\noindent
{\it Conclusions.}
In conclusion, the main achievement of this work is the development of a simple elasto-plastic model coupled to microscopic simulations via which all necessary parameters are inferred. This allows to simultaneously reproduce quantitatively the steady state flow properties, along with transient phenomena for a wide range of imposed strain rate and imposed stress, 
for different  preparation histories. We also match, in a qualitative manner, the steady state distributions of local yield thresholds.  In this process, we have illustrated how to identify and directly link coarse grained quantities from microscopic models, such as measurements of local stress and yield stress maps, that are used as initial inputs for elasto-plastic lattice models. Thereby, our work paves the road towards more quantitative and predictive multi-scale modeling in this field.

{\em Acknowledgements --} This work has been funded via the CEFIPRA Project 5604-1. We thank Francesco Puosi and Vishwas Vasisht for providing us with codes regarding the frozen matrix method. We thank Sylvain Patinet for sharing data and valuable discussions. All computations were mainly performed at Nandadevi HPC cluster of IMSc and on the Froggy platform of the 
\href{https://ciment.ujf-grenoble.fr}{CIMENT} infrastructure supported by the Rh\^one-Alpes region (GRANT CPER07-13
\href{http://www.ci-ra.org/}{CIRA}) and the Equip@Meso project (reference ANR-10-EQPX-29-01).

{\em Contributions.} Chen Liu and Suman Dutta contributed equally to this work.

\bibliography{apssamp}

\providecommand{\noopsort}[1]{}\providecommand{\singleletter}[1]{#1}%
\begin{thebibliography}{60}%
\makeatletter
\providecommand \@ifxundefined [1]{%
 \@ifx{#1\undefined}
}%
\providecommand \@ifnum [1]{%
 \ifnum #1\expandafter \@firstoftwo
 \else \expandafter \@secondoftwo
 \fi
}%
\providecommand \@ifx [1]{%
 \ifx #1\expandafter \@firstoftwo
 \else \expandafter \@secondoftwo
 \fi
}%
\providecommand \natexlab [1]{#1}%
\providecommand \enquote  [1]{``#1''}%
\providecommand \bibnamefont  [1]{#1}%
\providecommand \bibfnamefont [1]{#1}%
\providecommand \citenamefont [1]{#1}%
\providecommand \href@noop [0]{\@secondoftwo}%
\providecommand \href [0]{\begingroup \@sanitize@url \@href}%
\providecommand \@href[1]{\@@startlink{#1}\@@href}%
\providecommand \@@href[1]{\endgroup#1\@@endlink}%
\providecommand \@sanitize@url [0]{\catcode `\\12\catcode `\$12\catcode
  `\&12\catcode `\#12\catcode `\^12\catcode `\_12\catcode `\%12\relax}%
\providecommand \@@startlink[1]{}%
\providecommand \@@endlink[0]{}%
\providecommand \url  [0]{\begingroup\@sanitize@url \@url }%
\providecommand \@url [1]{\endgroup\@href {#1}{\urlprefix }}%
\providecommand \urlprefix  [0]{URL }%
\providecommand \Eprint [0]{\href }%
\providecommand \doibase [0]{https://doi.org/}%
\providecommand \selectlanguage [0]{\@gobble}%
\providecommand \bibinfo  [0]{\@secondoftwo}%
\providecommand \bibfield  [0]{\@secondoftwo}%
\providecommand \translation [1]{[#1]}%
\providecommand \BibitemOpen [0]{}%
\providecommand \bibitemStop [0]{}%
\providecommand \bibitemNoStop [0]{.\EOS\space}%
\providecommand \EOS [0]{\spacefactor3000\relax}%
\providecommand \BibitemShut  [1]{\csname bibitem#1\endcsname}%
\let\auto@bib@innerbib\@empty
\bibitem [{\citenamefont {Rodney}\ \emph {et~al.}(2011)\citenamefont {Rodney},
  \citenamefont {Tanguy},\ and\ \citenamefont
  {Vandembroucq}}]{rodney2011modeling}%
  \BibitemOpen
  \bibfield  {author} {\bibinfo {author} {\bibfnamefont {D.}~\bibnamefont
  {Rodney}}, \bibinfo {author} {\bibfnamefont {A.}~\bibnamefont {Tanguy}},\
  and\ \bibinfo {author} {\bibfnamefont {D.}~\bibnamefont {Vandembroucq}},\
  }\bibfield  {title} {\bibinfo {title} {Modeling the mechanics of amorphous
  solids at different length scale and time scale},\ }\href@noop {} {\bibfield
  {journal} {\bibinfo  {journal} {Modelling and Simulation in Materials Science
  and Engineering}\ }\textbf {\bibinfo {volume} {19}},\ \bibinfo {pages}
  {083001} (\bibinfo {year} {2011})}\BibitemShut {NoStop}%
\bibitem [{\citenamefont {Bonn}\ \emph {et~al.}(2017)\citenamefont {Bonn},
  \citenamefont {Denn}, \citenamefont {Berthier}, \citenamefont {Divoux},\ and\
  \citenamefont {Manneville}}]{bonn2017yield}%
  \BibitemOpen
  \bibfield  {author} {\bibinfo {author} {\bibfnamefont {D.}~\bibnamefont
  {Bonn}}, \bibinfo {author} {\bibfnamefont {M.~M.}\ \bibnamefont {Denn}},
  \bibinfo {author} {\bibfnamefont {L.}~\bibnamefont {Berthier}}, \bibinfo
  {author} {\bibfnamefont {T.}~\bibnamefont {Divoux}},\ and\ \bibinfo {author}
  {\bibfnamefont {S.}~\bibnamefont {Manneville}},\ }\bibfield  {title}
  {\bibinfo {title} {Yield stress materials in soft condensed matter},\
  }\href@noop {} {\bibfield  {journal} {\bibinfo  {journal} {Reviews of Modern
  Physics}\ }\textbf {\bibinfo {volume} {89}},\ \bibinfo {pages} {035005}
  (\bibinfo {year} {2017})}\BibitemShut {NoStop}%
\bibitem [{\citenamefont {Nicolas}\ \emph {et~al.}(2018)\citenamefont
  {Nicolas}, \citenamefont {Ferrero}, \citenamefont {Martens},\ and\
  \citenamefont {Barrat}}]{nicolas2018deformation}%
  \BibitemOpen
  \bibfield  {author} {\bibinfo {author} {\bibfnamefont {A.}~\bibnamefont
  {Nicolas}}, \bibinfo {author} {\bibfnamefont {E.~E.}\ \bibnamefont
  {Ferrero}}, \bibinfo {author} {\bibfnamefont {K.}~\bibnamefont {Martens}},\
  and\ \bibinfo {author} {\bibfnamefont {J.-L.}\ \bibnamefont {Barrat}},\
  }\bibfield  {title} {\bibinfo {title} {Deformation and flow of amorphous
  solids: Insights from elastoplastic models},\ }\href@noop {} {\bibfield
  {journal} {\bibinfo  {journal} {Reviews of Modern Physics}\ }\textbf
  {\bibinfo {volume} {90}},\ \bibinfo {pages} {045006} (\bibinfo {year}
  {2018})}\BibitemShut {NoStop}%
\bibitem [{\citenamefont {Cipelletti}\ \emph {et~al.}(2020)\citenamefont
  {Cipelletti}, \citenamefont {Martens},\ and\ \citenamefont
  {Ramos}}]{cipelletti2020microscopic}%
  \BibitemOpen
  \bibfield  {author} {\bibinfo {author} {\bibfnamefont {L.}~\bibnamefont
  {Cipelletti}}, \bibinfo {author} {\bibfnamefont {K.}~\bibnamefont
  {Martens}},\ and\ \bibinfo {author} {\bibfnamefont {L.}~\bibnamefont
  {Ramos}},\ }\bibfield  {title} {\bibinfo {title} {Microscopic precursors of
  failure in soft matter},\ }\href@noop {} {\bibfield  {journal} {\bibinfo
  {journal} {Soft matter}\ }\textbf {\bibinfo {volume} {16}},\ \bibinfo {pages}
  {82} (\bibinfo {year} {2020})}\BibitemShut {NoStop}%
\bibitem [{\citenamefont {Kawamura}\ \emph {et~al.}(1999)\citenamefont
  {Kawamura}, \citenamefont {Shibata}, \citenamefont {Inoue},\ and\
  \citenamefont {Masumoto}}]{kawamura1999stress}%
  \BibitemOpen
  \bibfield  {author} {\bibinfo {author} {\bibfnamefont {Y.}~\bibnamefont
  {Kawamura}}, \bibinfo {author} {\bibfnamefont {T.}~\bibnamefont {Shibata}},
  \bibinfo {author} {\bibfnamefont {A.}~\bibnamefont {Inoue}},\ and\ \bibinfo
  {author} {\bibfnamefont {T.}~\bibnamefont {Masumoto}},\ }\bibfield  {title}
  {\bibinfo {title} {Stress overshoot in stress-strain curves of
  zr65al10ni10cu15 metallic glass},\ }\href@noop {} {\bibfield  {journal}
  {\bibinfo  {journal} {Materials Transactions, JIM}\ }\textbf {\bibinfo
  {volume} {40}},\ \bibinfo {pages} {335} (\bibinfo {year} {1999})}\BibitemShut
  {NoStop}%
\bibitem [{\citenamefont {Lu}\ \emph {et~al.}(2003)\citenamefont {Lu},
  \citenamefont {Ravichandran},\ and\ \citenamefont
  {Johnson}}]{lu2003deformation}%
  \BibitemOpen
  \bibfield  {author} {\bibinfo {author} {\bibfnamefont {J.}~\bibnamefont
  {Lu}}, \bibinfo {author} {\bibfnamefont {G.}~\bibnamefont {Ravichandran}},\
  and\ \bibinfo {author} {\bibfnamefont {W.~L.}\ \bibnamefont {Johnson}},\
  }\bibfield  {title} {\bibinfo {title} {Deformation behavior of the zr41.
  2ti13. 8cu12. 5ni10be22. 5 bulk metallic glass over a wide range of
  strain-rates and temperatures},\ }\href@noop {} {\bibfield  {journal}
  {\bibinfo  {journal} {Acta materialia}\ }\textbf {\bibinfo {volume} {51}},\
  \bibinfo {pages} {3429} (\bibinfo {year} {2003})}\BibitemShut {NoStop}%
\bibitem [{\citenamefont {Maa{\ss}}\ \emph {et~al.}(2012)\citenamefont
  {Maa{\ss}}, \citenamefont {Klaum{\"u}nzer}, \citenamefont {Villard},
  \citenamefont {Derlet},\ and\ \citenamefont {L{\"o}ffler}}]{maass2012shear}%
  \BibitemOpen
  \bibfield  {author} {\bibinfo {author} {\bibfnamefont {R.}~\bibnamefont
  {Maa{\ss}}}, \bibinfo {author} {\bibfnamefont {D.}~\bibnamefont
  {Klaum{\"u}nzer}}, \bibinfo {author} {\bibfnamefont {G.}~\bibnamefont
  {Villard}}, \bibinfo {author} {\bibfnamefont {P.}~\bibnamefont {Derlet}},\
  and\ \bibinfo {author} {\bibfnamefont {J.~F.}\ \bibnamefont {L{\"o}ffler}},\
  }\bibfield  {title} {\bibinfo {title} {Shear-band arrest and stress
  overshoots during inhomogeneous flow in a metallic glass},\ }\href@noop {}
  {\bibfield  {journal} {\bibinfo  {journal} {Applied Physics Letters}\
  }\textbf {\bibinfo {volume} {100}},\ \bibinfo {pages} {071904} (\bibinfo
  {year} {2012})}\BibitemShut {NoStop}%
\bibitem [{\citenamefont {Amann}\ \emph {et~al.}(2013)\citenamefont {Amann},
  \citenamefont {Siebenb{\"u}rger}, \citenamefont {Kr{\"u}ger}, \citenamefont
  {Weysser}, \citenamefont {Ballauff},\ and\ \citenamefont
  {Fuchs}}]{amann2013overshoots}%
  \BibitemOpen
  \bibfield  {author} {\bibinfo {author} {\bibfnamefont {C.~P.}\ \bibnamefont
  {Amann}}, \bibinfo {author} {\bibfnamefont {M.}~\bibnamefont
  {Siebenb{\"u}rger}}, \bibinfo {author} {\bibfnamefont {M.}~\bibnamefont
  {Kr{\"u}ger}}, \bibinfo {author} {\bibfnamefont {F.}~\bibnamefont {Weysser}},
  \bibinfo {author} {\bibfnamefont {M.}~\bibnamefont {Ballauff}},\ and\
  \bibinfo {author} {\bibfnamefont {M.}~\bibnamefont {Fuchs}},\ }\bibfield
  {title} {\bibinfo {title} {Overshoots in stress-strain curves: Colloid
  experiments and schematic mode coupling theory},\ }\href@noop {} {\bibfield
  {journal} {\bibinfo  {journal} {Journal of Rheology}\ }\textbf {\bibinfo
  {volume} {57}},\ \bibinfo {pages} {149} (\bibinfo {year} {2013})}\BibitemShut
  {NoStop}%
\bibitem [{\citenamefont {Divoux}\ \emph {et~al.}(2011)\citenamefont {Divoux},
  \citenamefont {Barentin},\ and\ \citenamefont
  {Manneville}}]{divoux2011stress}%
  \BibitemOpen
  \bibfield  {author} {\bibinfo {author} {\bibfnamefont {T.}~\bibnamefont
  {Divoux}}, \bibinfo {author} {\bibfnamefont {C.}~\bibnamefont {Barentin}},\
  and\ \bibinfo {author} {\bibfnamefont {S.}~\bibnamefont {Manneville}},\
  }\bibfield  {title} {\bibinfo {title} {Stress overshoot in a simple yield
  stress fluid: An extensive study combining rheology and velocimetry},\
  }\href@noop {} {\bibfield  {journal} {\bibinfo  {journal} {Soft Matter}\
  }\textbf {\bibinfo {volume} {7}},\ \bibinfo {pages} {9335} (\bibinfo {year}
  {2011})}\BibitemShut {NoStop}%
\bibitem [{\citenamefont {Dzuy}\ and\ \citenamefont
  {Boger}(1983)}]{dzuy1983yield}%
  \BibitemOpen
  \bibfield  {author} {\bibinfo {author} {\bibfnamefont {N.~Q.}\ \bibnamefont
  {Dzuy}}\ and\ \bibinfo {author} {\bibfnamefont {D.~V.}\ \bibnamefont
  {Boger}},\ }\bibfield  {title} {\bibinfo {title} {Yield stress measurement
  for concentrated suspensions},\ }\href@noop {} {\bibfield  {journal}
  {\bibinfo  {journal} {Journal of Rheology}\ }\textbf {\bibinfo {volume}
  {27}},\ \bibinfo {pages} {321} (\bibinfo {year} {1983})}\BibitemShut
  {NoStop}%
\bibitem [{\citenamefont {Rogers}\ \emph {et~al.}(2010)\citenamefont {Rogers},
  \citenamefont {Callaghan}, \citenamefont {Petekidis},\ and\ \citenamefont
  {Vlassopoulos}}]{rogers2010time}%
  \BibitemOpen
  \bibfield  {author} {\bibinfo {author} {\bibfnamefont {S.}~\bibnamefont
  {Rogers}}, \bibinfo {author} {\bibfnamefont {P.}~\bibnamefont {Callaghan}},
  \bibinfo {author} {\bibfnamefont {G.}~\bibnamefont {Petekidis}},\ and\
  \bibinfo {author} {\bibfnamefont {D.}~\bibnamefont {Vlassopoulos}},\
  }\bibfield  {title} {\bibinfo {title} {Time-dependent rheology of colloidal
  star glasses},\ }\href@noop {} {\bibfield  {journal} {\bibinfo  {journal}
  {Journal of Rheology}\ }\textbf {\bibinfo {volume} {54}},\ \bibinfo {pages}
  {133} (\bibinfo {year} {2010})}\BibitemShut {NoStop}%
\bibitem [{\citenamefont {Zausch}\ \emph {et~al.}(2008)\citenamefont {Zausch},
  \citenamefont {Horbach}, \citenamefont {Laurati}, \citenamefont {Egelhaaf},
  \citenamefont {Brader}, \citenamefont {Voigtmann},\ and\ \citenamefont
  {Fuchs}}]{zausch2008equilibrium}%
  \BibitemOpen
  \bibfield  {author} {\bibinfo {author} {\bibfnamefont {J.}~\bibnamefont
  {Zausch}}, \bibinfo {author} {\bibfnamefont {J.}~\bibnamefont {Horbach}},
  \bibinfo {author} {\bibfnamefont {M.}~\bibnamefont {Laurati}}, \bibinfo
  {author} {\bibfnamefont {S.~U.}\ \bibnamefont {Egelhaaf}}, \bibinfo {author}
  {\bibfnamefont {J.~M.}\ \bibnamefont {Brader}}, \bibinfo {author}
  {\bibfnamefont {T.}~\bibnamefont {Voigtmann}},\ and\ \bibinfo {author}
  {\bibfnamefont {M.}~\bibnamefont {Fuchs}},\ }\bibfield  {title} {\bibinfo
  {title} {From equilibrium to steady state: the transient dynamics of
  colloidal liquids under shear},\ }\href@noop {} {\bibfield  {journal}
  {\bibinfo  {journal} {Journal of Physics: Condensed Matter}\ }\textbf
  {\bibinfo {volume} {20}},\ \bibinfo {pages} {404210} (\bibinfo {year}
  {2008})}\BibitemShut {NoStop}%
\bibitem [{\citenamefont {Sentjabrskaja}\ \emph {et~al.}(2014)\citenamefont
  {Sentjabrskaja}, \citenamefont {Hermes}, \citenamefont {Poon}, \citenamefont
  {Estrada}, \citenamefont {Castaneda-Priego}, \citenamefont {Egelhaaf},\ and\
  \citenamefont {Laurati}}]{sentjabrskaja2014transient}%
  \BibitemOpen
  \bibfield  {author} {\bibinfo {author} {\bibfnamefont {T.}~\bibnamefont
  {Sentjabrskaja}}, \bibinfo {author} {\bibfnamefont {M.}~\bibnamefont
  {Hermes}}, \bibinfo {author} {\bibfnamefont {W.}~\bibnamefont {Poon}},
  \bibinfo {author} {\bibfnamefont {C.}~\bibnamefont {Estrada}}, \bibinfo
  {author} {\bibfnamefont {R.}~\bibnamefont {Castaneda-Priego}}, \bibinfo
  {author} {\bibfnamefont {S.}~\bibnamefont {Egelhaaf}},\ and\ \bibinfo
  {author} {\bibfnamefont {M.}~\bibnamefont {Laurati}},\ }\bibfield  {title}
  {\bibinfo {title} {Transient dynamics during stress overshoots in binary
  colloidal glasses},\ }\href@noop {} {\bibfield  {journal} {\bibinfo
  {journal} {Soft Matter}\ }\textbf {\bibinfo {volume} {10}},\ \bibinfo {pages}
  {6546} (\bibinfo {year} {2014})}\BibitemShut {NoStop}%
\bibitem [{\citenamefont {Koumakis}\ \emph
  {et~al.}(2012{\natexlab{a}})\citenamefont {Koumakis}, \citenamefont
  {Laurati}, \citenamefont {Egelhaaf}, \citenamefont {Brady},\ and\
  \citenamefont {Petekidis}}]{koumakis2012yielding}%
  \BibitemOpen
  \bibfield  {author} {\bibinfo {author} {\bibfnamefont {N.}~\bibnamefont
  {Koumakis}}, \bibinfo {author} {\bibfnamefont {M.}~\bibnamefont {Laurati}},
  \bibinfo {author} {\bibfnamefont {S.}~\bibnamefont {Egelhaaf}}, \bibinfo
  {author} {\bibfnamefont {J.}~\bibnamefont {Brady}},\ and\ \bibinfo {author}
  {\bibfnamefont {G.}~\bibnamefont {Petekidis}},\ }\bibfield  {title} {\bibinfo
  {title} {Yielding of hard-sphere glasses during start-up shear},\ }\href@noop
  {} {\bibfield  {journal} {\bibinfo  {journal} {Physical review letters}\
  }\textbf {\bibinfo {volume} {108}},\ \bibinfo {pages} {098303} (\bibinfo
  {year} {2012}{\natexlab{a}})}\BibitemShut {NoStop}%
\bibitem [{\citenamefont {Koumakis}\ \emph
  {et~al.}(2012{\natexlab{b}})\citenamefont {Koumakis}, \citenamefont
  {Pamvouxoglou}, \citenamefont {Poulos},\ and\ \citenamefont
  {Petekidis}}]{koumakis2012direct}%
  \BibitemOpen
  \bibfield  {author} {\bibinfo {author} {\bibfnamefont {N.}~\bibnamefont
  {Koumakis}}, \bibinfo {author} {\bibfnamefont {A.}~\bibnamefont
  {Pamvouxoglou}}, \bibinfo {author} {\bibfnamefont {A.}~\bibnamefont
  {Poulos}},\ and\ \bibinfo {author} {\bibfnamefont {G.}~\bibnamefont
  {Petekidis}},\ }\bibfield  {title} {\bibinfo {title} {Direct comparison of
  the rheology of model hard and soft particle glasses},\ }\href@noop {}
  {\bibfield  {journal} {\bibinfo  {journal} {Soft Matter}\ }\textbf {\bibinfo
  {volume} {8}},\ \bibinfo {pages} {4271} (\bibinfo {year}
  {2012}{\natexlab{b}})}\BibitemShut {NoStop}%
\bibitem [{\citenamefont {Chaudhuri}\ and\ \citenamefont
  {Horbach}(2013)}]{chaudhuri2013onset}%
  \BibitemOpen
  \bibfield  {author} {\bibinfo {author} {\bibfnamefont {P.}~\bibnamefont
  {Chaudhuri}}\ and\ \bibinfo {author} {\bibfnamefont {J.}~\bibnamefont
  {Horbach}},\ }\bibfield  {title} {\bibinfo {title} {Onset of flow in a
  confined colloidal glass under an imposed shear stress},\ }\href@noop {}
  {\bibfield  {journal} {\bibinfo  {journal} {Physical Review E}\ }\textbf
  {\bibinfo {volume} {88}},\ \bibinfo {pages} {040301} (\bibinfo {year}
  {2013})}\BibitemShut {NoStop}%
\bibitem [{\citenamefont {Sentjabrskaja}\ \emph {et~al.}(2015)\citenamefont
  {Sentjabrskaja}, \citenamefont {Chaudhuri}, \citenamefont {Hermes},
  \citenamefont {Poon}, \citenamefont {Horbach}, \citenamefont {Egelhaaf},\
  and\ \citenamefont {Laurati}}]{sentjabrskaja2015creep}%
  \BibitemOpen
  \bibfield  {author} {\bibinfo {author} {\bibfnamefont {T.}~\bibnamefont
  {Sentjabrskaja}}, \bibinfo {author} {\bibfnamefont {P.}~\bibnamefont
  {Chaudhuri}}, \bibinfo {author} {\bibfnamefont {M.}~\bibnamefont {Hermes}},
  \bibinfo {author} {\bibfnamefont {W.}~\bibnamefont {Poon}}, \bibinfo {author}
  {\bibfnamefont {J.}~\bibnamefont {Horbach}}, \bibinfo {author} {\bibfnamefont
  {S.}~\bibnamefont {Egelhaaf}},\ and\ \bibinfo {author} {\bibfnamefont
  {M.}~\bibnamefont {Laurati}},\ }\bibfield  {title} {\bibinfo {title} {Creep
  and flow of glasses: Strain response linked to the spatial distribution of
  dynamical heterogeneities},\ }\href@noop {} {\bibfield  {journal} {\bibinfo
  {journal} {Scientific reports}\ }\textbf {\bibinfo {volume} {5}},\ \bibinfo
  {pages} {11884} (\bibinfo {year} {2015})}\BibitemShut {NoStop}%
\bibitem [{\citenamefont {Agoritsas}\ and\ \citenamefont
  {Martens}(2017)}]{agoritsas2017non}%
  \BibitemOpen
  \bibfield  {author} {\bibinfo {author} {\bibfnamefont {E.}~\bibnamefont
  {Agoritsas}}\ and\ \bibinfo {author} {\bibfnamefont {K.}~\bibnamefont
  {Martens}},\ }\bibfield  {title} {\bibinfo {title} {Non-trivial rheological
  exponents in sheared yield stress fluids},\ }\href@noop {} {\bibfield
  {journal} {\bibinfo  {journal} {Soft Matter}\ }\textbf {\bibinfo {volume}
  {13}},\ \bibinfo {pages} {4653} (\bibinfo {year} {2017})}\BibitemShut
  {NoStop}%
\bibitem [{\citenamefont {Argon}(1979)}]{argon1979plastic}%
  \BibitemOpen
  \bibfield  {author} {\bibinfo {author} {\bibfnamefont {A.}~\bibnamefont
  {Argon}},\ }\bibfield  {title} {\bibinfo {title} {Plastic deformation in
  metallic glasses},\ }\href@noop {} {\bibfield  {journal} {\bibinfo  {journal}
  {Acta metall.}\ }\textbf {\bibinfo {volume} {27}},\ \bibinfo {pages} {47}
  (\bibinfo {year} {1979})}\BibitemShut {NoStop}%
\bibitem [{\citenamefont {Eshelby}(1957)}]{eshelby1957proc}%
  \BibitemOpen
  \bibfield  {author} {\bibinfo {author} {\bibfnamefont {J.}~\bibnamefont
  {Eshelby}},\ }\bibfield  {title} {\bibinfo {title} {Proc. r. soc. london,
  ser. a},\ }\href@noop {} {\  (\bibinfo {year} {1957})}\BibitemShut {NoStop}%
\bibitem [{\citenamefont {Baret}\ \emph {et~al.}(2002)\citenamefont {Baret},
  \citenamefont {Vandembroucq},\ and\ \citenamefont
  {Roux}}]{baret2002extremal}%
  \BibitemOpen
  \bibfield  {author} {\bibinfo {author} {\bibfnamefont {J.-C.}\ \bibnamefont
  {Baret}}, \bibinfo {author} {\bibfnamefont {D.}~\bibnamefont
  {Vandembroucq}},\ and\ \bibinfo {author} {\bibfnamefont {S.}~\bibnamefont
  {Roux}},\ }\bibfield  {title} {\bibinfo {title} {Extremal model for amorphous
  media plasticity},\ }\href@noop {} {\bibfield  {journal} {\bibinfo  {journal}
  {Physical review letters}\ }\textbf {\bibinfo {volume} {89}},\ \bibinfo
  {pages} {195506} (\bibinfo {year} {2002})}\BibitemShut {NoStop}%
\bibitem [{\citenamefont {Martens}\ \emph {et~al.}(2011)\citenamefont
  {Martens}, \citenamefont {Bocquet},\ and\ \citenamefont
  {Barrat}}]{martens2011connecting}%
  \BibitemOpen
  \bibfield  {author} {\bibinfo {author} {\bibfnamefont {K.}~\bibnamefont
  {Martens}}, \bibinfo {author} {\bibfnamefont {L.}~\bibnamefont {Bocquet}},\
  and\ \bibinfo {author} {\bibfnamefont {J.-L.}\ \bibnamefont {Barrat}},\
  }\bibfield  {title} {\bibinfo {title} {Connecting diffusion and dynamical
  heterogeneities in actively deformed amorphous systems},\ }\href@noop {}
  {\bibfield  {journal} {\bibinfo  {journal} {Physical review letters}\
  }\textbf {\bibinfo {volume} {106}},\ \bibinfo {pages} {156001} (\bibinfo
  {year} {2011})}\BibitemShut {NoStop}%
\bibitem [{\citenamefont {Tsamados}\ \emph {et~al.}(2009)\citenamefont
  {Tsamados}, \citenamefont {Tanguy}, \citenamefont {Goldenberg},\ and\
  \citenamefont {Barrat}}]{tsamados2009local}%
  \BibitemOpen
  \bibfield  {author} {\bibinfo {author} {\bibfnamefont {M.}~\bibnamefont
  {Tsamados}}, \bibinfo {author} {\bibfnamefont {A.}~\bibnamefont {Tanguy}},
  \bibinfo {author} {\bibfnamefont {C.}~\bibnamefont {Goldenberg}},\ and\
  \bibinfo {author} {\bibfnamefont {J.-L.}\ \bibnamefont {Barrat}},\ }\bibfield
   {title} {\bibinfo {title} {Local elasticity map and plasticity in a model
  lennard-jones glass},\ }\href@noop {} {\bibfield  {journal} {\bibinfo
  {journal} {Physical Review E}\ }\textbf {\bibinfo {volume} {80}},\ \bibinfo
  {pages} {026112} (\bibinfo {year} {2009})}\BibitemShut {NoStop}%
\bibitem [{\citenamefont {Sollich}\ \emph {et~al.}(1997)\citenamefont
  {Sollich}, \citenamefont {Lequeux}, \citenamefont {H{\'e}braud},\ and\
  \citenamefont {Cates}}]{sollich1997rheology}%
  \BibitemOpen
  \bibfield  {author} {\bibinfo {author} {\bibfnamefont {P.}~\bibnamefont
  {Sollich}}, \bibinfo {author} {\bibfnamefont {F.}~\bibnamefont {Lequeux}},
  \bibinfo {author} {\bibfnamefont {P.}~\bibnamefont {H{\'e}braud}},\ and\
  \bibinfo {author} {\bibfnamefont {M.~E.}\ \bibnamefont {Cates}},\ }\bibfield
  {title} {\bibinfo {title} {Rheology of soft glassy materials},\ }\href@noop
  {} {\bibfield  {journal} {\bibinfo  {journal} {Physical review letters}\
  }\textbf {\bibinfo {volume} {78}},\ \bibinfo {pages} {2020} (\bibinfo {year}
  {1997})}\BibitemShut {NoStop}%
\bibitem [{\citenamefont {Moorcroft}\ \emph {et~al.}(2011)\citenamefont
  {Moorcroft}, \citenamefont {Cates},\ and\ \citenamefont
  {Fielding}}]{moorcroft2011age}%
  \BibitemOpen
  \bibfield  {author} {\bibinfo {author} {\bibfnamefont {R.~L.}\ \bibnamefont
  {Moorcroft}}, \bibinfo {author} {\bibfnamefont {M.~E.}\ \bibnamefont
  {Cates}},\ and\ \bibinfo {author} {\bibfnamefont {S.~M.}\ \bibnamefont
  {Fielding}},\ }\bibfield  {title} {\bibinfo {title} {Age-dependent transient
  shear banding in soft glasses},\ }\href@noop {} {\bibfield  {journal}
  {\bibinfo  {journal} {Physical review letters}\ }\textbf {\bibinfo {volume}
  {106}},\ \bibinfo {pages} {055502} (\bibinfo {year} {2011})}\BibitemShut
  {NoStop}%
\bibitem [{\citenamefont {Picard}\ \emph {et~al.}(2002)\citenamefont {Picard},
  \citenamefont {Ajdari}, \citenamefont {Bocquet},\ and\ \citenamefont
  {Lequeux}}]{picard2002simple}%
  \BibitemOpen
  \bibfield  {author} {\bibinfo {author} {\bibfnamefont {G.}~\bibnamefont
  {Picard}}, \bibinfo {author} {\bibfnamefont {A.}~\bibnamefont {Ajdari}},
  \bibinfo {author} {\bibfnamefont {L.}~\bibnamefont {Bocquet}},\ and\ \bibinfo
  {author} {\bibfnamefont {F.}~\bibnamefont {Lequeux}},\ }\bibfield  {title}
  {\bibinfo {title} {Simple model for heterogeneous flows of yield stress
  fluids},\ }\href@noop {} {\bibfield  {journal} {\bibinfo  {journal} {Physical
  Review E}\ }\textbf {\bibinfo {volume} {66}},\ \bibinfo {pages} {051501}
  (\bibinfo {year} {2002})}\BibitemShut {NoStop}%
\bibitem [{\citenamefont {Bocquet}\ \emph {et~al.}(2009)\citenamefont
  {Bocquet}, \citenamefont {Colin},\ and\ \citenamefont
  {Ajdari}}]{bocquet2009kinetic}%
  \BibitemOpen
  \bibfield  {author} {\bibinfo {author} {\bibfnamefont {L.}~\bibnamefont
  {Bocquet}}, \bibinfo {author} {\bibfnamefont {A.}~\bibnamefont {Colin}},\
  and\ \bibinfo {author} {\bibfnamefont {A.}~\bibnamefont {Ajdari}},\
  }\bibfield  {title} {\bibinfo {title} {Kinetic theory of plastic flow in soft
  glassy materials},\ }\href@noop {} {\bibfield  {journal} {\bibinfo  {journal}
  {Physical review letters}\ }\textbf {\bibinfo {volume} {103}},\ \bibinfo
  {pages} {036001} (\bibinfo {year} {2009})}\BibitemShut {NoStop}%
\bibitem [{\citenamefont {Falk}\ and\ \citenamefont
  {Langer}(1998)}]{falk1998dynamics}%
  \BibitemOpen
  \bibfield  {author} {\bibinfo {author} {\bibfnamefont {M.~L.}\ \bibnamefont
  {Falk}}\ and\ \bibinfo {author} {\bibfnamefont {J.~S.}\ \bibnamefont
  {Langer}},\ }\bibfield  {title} {\bibinfo {title} {Dynamics of viscoplastic
  deformation in amorphous solids},\ }\href@noop {} {\bibfield  {journal}
  {\bibinfo  {journal} {Physical Review E}\ }\textbf {\bibinfo {volume} {57}},\
  \bibinfo {pages} {7192} (\bibinfo {year} {1998})}\BibitemShut {NoStop}%
\bibitem [{\citenamefont {Fuchs}\ and\ \citenamefont
  {Cates}(2002)}]{fuchs2002theory}%
  \BibitemOpen
  \bibfield  {author} {\bibinfo {author} {\bibfnamefont {M.}~\bibnamefont
  {Fuchs}}\ and\ \bibinfo {author} {\bibfnamefont {M.~E.}\ \bibnamefont
  {Cates}},\ }\bibfield  {title} {\bibinfo {title} {Theory of nonlinear
  rheology and yielding of dense colloidal suspensions},\ }\href@noop {}
  {\bibfield  {journal} {\bibinfo  {journal} {Physical review letters}\
  }\textbf {\bibinfo {volume} {89}},\ \bibinfo {pages} {248304} (\bibinfo
  {year} {2002})}\BibitemShut {NoStop}%
\bibitem [{\citenamefont {H{\'e}braud}\ and\ \citenamefont
  {Lequeux}(1998)}]{hebraud1998mode}%
  \BibitemOpen
  \bibfield  {author} {\bibinfo {author} {\bibfnamefont {P.}~\bibnamefont
  {H{\'e}braud}}\ and\ \bibinfo {author} {\bibfnamefont {F.}~\bibnamefont
  {Lequeux}},\ }\bibfield  {title} {\bibinfo {title} {Mode-coupling theory for
  the pasty rheology of soft glassy materials},\ }\href@noop {} {\bibfield
  {journal} {\bibinfo  {journal} {Physical review letters}\ }\textbf {\bibinfo
  {volume} {81}},\ \bibinfo {pages} {2934} (\bibinfo {year}
  {1998})}\BibitemShut {NoStop}%
\bibitem [{\citenamefont {Merabia}\ and\ \citenamefont
  {Detcheverry}(2016)}]{merabia2016thermally}%
  \BibitemOpen
  \bibfield  {author} {\bibinfo {author} {\bibfnamefont {S.}~\bibnamefont
  {Merabia}}\ and\ \bibinfo {author} {\bibfnamefont {F.}~\bibnamefont
  {Detcheverry}},\ }\bibfield  {title} {\bibinfo {title} {Thermally activated
  creep and fluidization in flowing disordered materials},\ }\href@noop {}
  {\bibfield  {journal} {\bibinfo  {journal} {EPL (Europhysics Letters)}\
  }\textbf {\bibinfo {volume} {116}},\ \bibinfo {pages} {46003} (\bibinfo
  {year} {2016})}\BibitemShut {NoStop}%
\bibitem [{\citenamefont {Manning}\ \emph {et~al.}(2007)\citenamefont
  {Manning}, \citenamefont {Langer},\ and\ \citenamefont
  {Carlson}}]{manning2007strain}%
  \BibitemOpen
  \bibfield  {author} {\bibinfo {author} {\bibfnamefont {M.~L.}\ \bibnamefont
  {Manning}}, \bibinfo {author} {\bibfnamefont {J.~S.}\ \bibnamefont
  {Langer}},\ and\ \bibinfo {author} {\bibfnamefont {J.}~\bibnamefont
  {Carlson}},\ }\bibfield  {title} {\bibinfo {title} {Strain localization in a
  shear transformation zone model for amorphous solids},\ }\href@noop {}
  {\bibfield  {journal} {\bibinfo  {journal} {Physical review E}\ }\textbf
  {\bibinfo {volume} {76}},\ \bibinfo {pages} {056106} (\bibinfo {year}
  {2007})}\BibitemShut {NoStop}%
\bibitem [{\citenamefont {Goyon}\ \emph {et~al.}(2008)\citenamefont {Goyon},
  \citenamefont {Colin}, \citenamefont {Ovarlez}, \citenamefont {Ajdari},\ and\
  \citenamefont {Bocquet}}]{goyon2008spatial}%
  \BibitemOpen
  \bibfield  {author} {\bibinfo {author} {\bibfnamefont {J.}~\bibnamefont
  {Goyon}}, \bibinfo {author} {\bibfnamefont {A.}~\bibnamefont {Colin}},
  \bibinfo {author} {\bibfnamefont {G.}~\bibnamefont {Ovarlez}}, \bibinfo
  {author} {\bibfnamefont {A.}~\bibnamefont {Ajdari}},\ and\ \bibinfo {author}
  {\bibfnamefont {L.}~\bibnamefont {Bocquet}},\ }\bibfield  {title} {\bibinfo
  {title} {Spatial cooperativity in soft glassy flows},\ }\href@noop {}
  {\bibfield  {journal} {\bibinfo  {journal} {Nature}\ }\textbf {\bibinfo
  {volume} {454}},\ \bibinfo {pages} {84} (\bibinfo {year} {2008})}\BibitemShut
  {NoStop}%
\bibitem [{\citenamefont {Mansard}\ \emph {et~al.}(2013)\citenamefont
  {Mansard}, \citenamefont {Colin}, \citenamefont {Chaudhuri},\ and\
  \citenamefont {Bocquet}}]{mansard2013molecular}%
  \BibitemOpen
  \bibfield  {author} {\bibinfo {author} {\bibfnamefont {V.}~\bibnamefont
  {Mansard}}, \bibinfo {author} {\bibfnamefont {A.}~\bibnamefont {Colin}},
  \bibinfo {author} {\bibfnamefont {P.}~\bibnamefont {Chaudhuri}},\ and\
  \bibinfo {author} {\bibfnamefont {L.}~\bibnamefont {Bocquet}},\ }\bibfield
  {title} {\bibinfo {title} {A molecular dynamics study of non-local effects in
  the flow of soft jammed particles},\ }\href@noop {} {\bibfield  {journal}
  {\bibinfo  {journal} {Soft matter}\ }\textbf {\bibinfo {volume} {9}},\
  \bibinfo {pages} {7489} (\bibinfo {year} {2013})}\BibitemShut {NoStop}%
\bibitem [{\citenamefont {Nicolas}\ and\ \citenamefont
  {Barrat}(2013)}]{nicolas2013spatial}%
  \BibitemOpen
  \bibfield  {author} {\bibinfo {author} {\bibfnamefont {A.}~\bibnamefont
  {Nicolas}}\ and\ \bibinfo {author} {\bibfnamefont {J.-L.}\ \bibnamefont
  {Barrat}},\ }\bibfield  {title} {\bibinfo {title} {Spatial cooperativity in
  microchannel flows of soft jammed materials: A mesoscopic approach},\
  }\href@noop {} {\bibfield  {journal} {\bibinfo  {journal} {Physical review
  letters}\ }\textbf {\bibinfo {volume} {110}},\ \bibinfo {pages} {138304}
  (\bibinfo {year} {2013})}\BibitemShut {NoStop}%
\bibitem [{\citenamefont {Hinkle}\ \emph {et~al.}(2017)\citenamefont {Hinkle},
  \citenamefont {Rycroft}, \citenamefont {Shields},\ and\ \citenamefont
  {Falk}}]{hinkle2017coarse}%
  \BibitemOpen
  \bibfield  {author} {\bibinfo {author} {\bibfnamefont {A.~R.}\ \bibnamefont
  {Hinkle}}, \bibinfo {author} {\bibfnamefont {C.~H.}\ \bibnamefont {Rycroft}},
  \bibinfo {author} {\bibfnamefont {M.~D.}\ \bibnamefont {Shields}},\ and\
  \bibinfo {author} {\bibfnamefont {M.~L.}\ \bibnamefont {Falk}},\ }\bibfield
  {title} {\bibinfo {title} {Coarse graining atomistic simulations of
  plastically deforming amorphous solids},\ }\href@noop {} {\bibfield
  {journal} {\bibinfo  {journal} {Physical Review E}\ }\textbf {\bibinfo
  {volume} {95}},\ \bibinfo {pages} {053001} (\bibinfo {year}
  {2017})}\BibitemShut {NoStop}%
\bibitem [{\citenamefont {Puosi}\ \emph {et~al.}(2014)\citenamefont {Puosi},
  \citenamefont {Rottler},\ and\ \citenamefont {Barrat}}]{puosi2014time}%
  \BibitemOpen
  \bibfield  {author} {\bibinfo {author} {\bibfnamefont {F.}~\bibnamefont
  {Puosi}}, \bibinfo {author} {\bibfnamefont {J.}~\bibnamefont {Rottler}},\
  and\ \bibinfo {author} {\bibfnamefont {J.-L.}\ \bibnamefont {Barrat}},\
  }\bibfield  {title} {\bibinfo {title} {Time-dependent elastic response to a
  local shear transformation in amorphous solids},\ }\href@noop {} {\bibfield
  {journal} {\bibinfo  {journal} {Physical Review E}\ }\textbf {\bibinfo
  {volume} {89}},\ \bibinfo {pages} {042302} (\bibinfo {year}
  {2014})}\BibitemShut {NoStop}%
\bibitem [{\citenamefont {Agoritsas}\ \emph {et~al.}(2015)\citenamefont
  {Agoritsas}, \citenamefont {Bertin}, \citenamefont {Martens},\ and\
  \citenamefont {Barrat}}]{agoritsas2015relevance}%
  \BibitemOpen
  \bibfield  {author} {\bibinfo {author} {\bibfnamefont {E.}~\bibnamefont
  {Agoritsas}}, \bibinfo {author} {\bibfnamefont {E.}~\bibnamefont {Bertin}},
  \bibinfo {author} {\bibfnamefont {K.}~\bibnamefont {Martens}},\ and\ \bibinfo
  {author} {\bibfnamefont {J.-L.}\ \bibnamefont {Barrat}},\ }\bibfield  {title}
  {\bibinfo {title} {On the relevance of disorder in athermal amorphous
  materials under shear},\ }\href@noop {} {\bibfield  {journal} {\bibinfo
  {journal} {The European Physical Journal E}\ }\textbf {\bibinfo {volume}
  {38}},\ \bibinfo {pages} {71} (\bibinfo {year} {2015})}\BibitemShut {NoStop}%
\bibitem [{\citenamefont {Nicolas}\ \emph {et~al.}(2014)\citenamefont
  {Nicolas}, \citenamefont {Rottler},\ and\ \citenamefont
  {Barrat}}]{nicolas2014spatiotemporal}%
  \BibitemOpen
  \bibfield  {author} {\bibinfo {author} {\bibfnamefont {A.}~\bibnamefont
  {Nicolas}}, \bibinfo {author} {\bibfnamefont {J.}~\bibnamefont {Rottler}},\
  and\ \bibinfo {author} {\bibfnamefont {J.-L.}\ \bibnamefont {Barrat}},\
  }\bibfield  {title} {\bibinfo {title} {Spatiotemporal correlations between
  plastic events in the shear flow of athermal amorphous solids},\ }\href@noop
  {} {\bibfield  {journal} {\bibinfo  {journal} {The European Physical Journal
  E}\ }\textbf {\bibinfo {volume} {37}},\ \bibinfo {pages} {50} (\bibinfo
  {year} {2014})}\BibitemShut {NoStop}%
\bibitem [{\citenamefont {Albaret}\ \emph {et~al.}(2016)\citenamefont
  {Albaret}, \citenamefont {Tanguy}, \citenamefont {Boioli},\ and\
  \citenamefont {Rodney}}]{albaret2016mapping}%
  \BibitemOpen
  \bibfield  {author} {\bibinfo {author} {\bibfnamefont {T.}~\bibnamefont
  {Albaret}}, \bibinfo {author} {\bibfnamefont {A.}~\bibnamefont {Tanguy}},
  \bibinfo {author} {\bibfnamefont {F.}~\bibnamefont {Boioli}},\ and\ \bibinfo
  {author} {\bibfnamefont {D.}~\bibnamefont {Rodney}},\ }\bibfield  {title}
  {\bibinfo {title} {Mapping between atomistic simulations and eshelby
  inclusions in the shear deformation of an amorphous silicon model},\
  }\href@noop {} {\bibfield  {journal} {\bibinfo  {journal} {Physical Review
  E}\ }\textbf {\bibinfo {volume} {93}},\ \bibinfo {pages} {053002} (\bibinfo
  {year} {2016})}\BibitemShut {NoStop}%
\bibitem [{\citenamefont {Boioli}\ \emph {et~al.}(2017)\citenamefont {Boioli},
  \citenamefont {Albaret},\ and\ \citenamefont {Rodney}}]{boioli2017shear}%
  \BibitemOpen
  \bibfield  {author} {\bibinfo {author} {\bibfnamefont {F.}~\bibnamefont
  {Boioli}}, \bibinfo {author} {\bibfnamefont {T.}~\bibnamefont {Albaret}},\
  and\ \bibinfo {author} {\bibfnamefont {D.}~\bibnamefont {Rodney}},\
  }\bibfield  {title} {\bibinfo {title} {Shear transformation distribution and
  activation in glasses at the atomic scale},\ }\href@noop {} {\bibfield
  {journal} {\bibinfo  {journal} {Physical Review E}\ }\textbf {\bibinfo
  {volume} {95}},\ \bibinfo {pages} {033005} (\bibinfo {year}
  {2017})}\BibitemShut {NoStop}%
\bibitem [{\citenamefont {Puosi}\ \emph {et~al.}(2015)\citenamefont {Puosi},
  \citenamefont {Olivier},\ and\ \citenamefont {Martens}}]{puosi2015probing}%
  \BibitemOpen
  \bibfield  {author} {\bibinfo {author} {\bibfnamefont {F.}~\bibnamefont
  {Puosi}}, \bibinfo {author} {\bibfnamefont {J.}~\bibnamefont {Olivier}},\
  and\ \bibinfo {author} {\bibfnamefont {K.}~\bibnamefont {Martens}},\
  }\bibfield  {title} {\bibinfo {title} {Probing relevant ingredients in
  mean-field approaches for the athermal rheology of yield stress materials},\
  }\href@noop {} {\bibfield  {journal} {\bibinfo  {journal} {Soft matter}\
  }\textbf {\bibinfo {volume} {11}},\ \bibinfo {pages} {7639} (\bibinfo {year}
  {2015})}\BibitemShut {NoStop}%
\bibitem [{\citenamefont {Patinet}\ \emph {et~al.}(2016)\citenamefont
  {Patinet}, \citenamefont {Vandembroucq},\ and\ \citenamefont
  {Falk}}]{patinet2016connecting}%
  \BibitemOpen
  \bibfield  {author} {\bibinfo {author} {\bibfnamefont {S.}~\bibnamefont
  {Patinet}}, \bibinfo {author} {\bibfnamefont {D.}~\bibnamefont
  {Vandembroucq}},\ and\ \bibinfo {author} {\bibfnamefont {M.~L.}\ \bibnamefont
  {Falk}},\ }\bibfield  {title} {\bibinfo {title} {Connecting local yield
  stresses with plastic activity in amorphous solids},\ }\href@noop {}
  {\bibfield  {journal} {\bibinfo  {journal} {Physical review letters}\
  }\textbf {\bibinfo {volume} {117}},\ \bibinfo {pages} {045501} (\bibinfo
  {year} {2016})}\BibitemShut {NoStop}%
\bibitem [{\citenamefont {Barbot}\ \emph {et~al.}(2018)\citenamefont {Barbot},
  \citenamefont {Lerbinger}, \citenamefont {Hernandez-Garcia}, \citenamefont
  {Garc{\'\i}a-Garc{\'\i}a}, \citenamefont {Falk}, \citenamefont
  {Vandembroucq},\ and\ \citenamefont {Patinet}}]{barbot2018local}%
  \BibitemOpen
  \bibfield  {author} {\bibinfo {author} {\bibfnamefont {A.}~\bibnamefont
  {Barbot}}, \bibinfo {author} {\bibfnamefont {M.}~\bibnamefont {Lerbinger}},
  \bibinfo {author} {\bibfnamefont {A.}~\bibnamefont {Hernandez-Garcia}},
  \bibinfo {author} {\bibfnamefont {R.}~\bibnamefont
  {Garc{\'\i}a-Garc{\'\i}a}}, \bibinfo {author} {\bibfnamefont {M.~L.}\
  \bibnamefont {Falk}}, \bibinfo {author} {\bibfnamefont {D.}~\bibnamefont
  {Vandembroucq}},\ and\ \bibinfo {author} {\bibfnamefont {S.}~\bibnamefont
  {Patinet}},\ }\bibfield  {title} {\bibinfo {title} {Local yield stress
  statistics in model amorphous solids},\ }\href@noop {} {\bibfield  {journal}
  {\bibinfo  {journal} {Physical Review E}\ }\textbf {\bibinfo {volume} {97}},\
  \bibinfo {pages} {033001} (\bibinfo {year} {2018})}\BibitemShut {NoStop}%
\bibitem [{\citenamefont {Barbot}\ \emph {et~al.}(2020)\citenamefont {Barbot},
  \citenamefont {Lerbinger}, \citenamefont {Lemaitre}, \citenamefont
  {Vandembroucq},\ and\ \citenamefont {Patinet}}]{barbot2020rejuvenation}%
  \BibitemOpen
  \bibfield  {author} {\bibinfo {author} {\bibfnamefont {A.}~\bibnamefont
  {Barbot}}, \bibinfo {author} {\bibfnamefont {M.}~\bibnamefont {Lerbinger}},
  \bibinfo {author} {\bibfnamefont {A.}~\bibnamefont {Lemaitre}}, \bibinfo
  {author} {\bibfnamefont {D.}~\bibnamefont {Vandembroucq}},\ and\ \bibinfo
  {author} {\bibfnamefont {S.}~\bibnamefont {Patinet}},\ }\bibfield  {title}
  {\bibinfo {title} {Rejuvenation and shear banding in model amorphous
  solids},\ }\href@noop {} {\bibfield  {journal} {\bibinfo  {journal} {Physical
  Review E}\ }\textbf {\bibinfo {volume} {101}},\ \bibinfo {pages} {033001}
  (\bibinfo {year} {2020})}\BibitemShut {NoStop}%
\bibitem [{\citenamefont {Lan{\c{c}}on}\ and\ \citenamefont
  {Billard}(1988)}]{lanccon1988two}%
  \BibitemOpen
  \bibfield  {author} {\bibinfo {author} {\bibfnamefont {F.}~\bibnamefont
  {Lan{\c{c}}on}}\ and\ \bibinfo {author} {\bibfnamefont {L.}~\bibnamefont
  {Billard}},\ }\bibfield  {title} {\bibinfo {title} {Two-dimensional system
  with a quasi-crystalline ground state},\ }\href@noop {} {\bibfield  {journal}
  {\bibinfo  {journal} {Journal de Physique}\ }\textbf {\bibinfo {volume}
  {49}},\ \bibinfo {pages} {249} (\bibinfo {year} {1988})}\BibitemShut
  {NoStop}%
\bibitem [{\citenamefont {Lan{\c{c}}on}\ \emph {et~al.}(1986)\citenamefont
  {Lan{\c{c}}on}, \citenamefont {Billard},\ and\ \citenamefont
  {Chaudhari}}]{lanccon1986thermodynamical}%
  \BibitemOpen
  \bibfield  {author} {\bibinfo {author} {\bibfnamefont {F.}~\bibnamefont
  {Lan{\c{c}}on}}, \bibinfo {author} {\bibfnamefont {L.}~\bibnamefont
  {Billard}},\ and\ \bibinfo {author} {\bibfnamefont {P.}~\bibnamefont
  {Chaudhari}},\ }\bibfield  {title} {\bibinfo {title} {Thermodynamical
  properties of a two-dimensional quasi-crystal from molecular dynamics
  calculations},\ }\href@noop {} {\bibfield  {journal} {\bibinfo  {journal}
  {EPL (Europhysics Letters)}\ }\textbf {\bibinfo {volume} {2}},\ \bibinfo
  {pages} {625} (\bibinfo {year} {1986})}\BibitemShut {NoStop}%
\bibitem [{si()}]{si}%
  \BibitemOpen
  \href@noop {} {\emph {\bibinfo {title} {See Supplemental Material at [URL
  will be inserted by publisher] for details of parameter tuning and our
  particle based simulations}}}\BibitemShut {NoStop}%
\bibitem [{\citenamefont {Sastry}\ \emph {et~al.}(1998)\citenamefont {Sastry},
  \citenamefont {Debenedetti},\ and\ \citenamefont
  {Stillinger}}]{sastry1998signatures}%
  \BibitemOpen
  \bibfield  {author} {\bibinfo {author} {\bibfnamefont {S.}~\bibnamefont
  {Sastry}}, \bibinfo {author} {\bibfnamefont {P.~G.}\ \bibnamefont
  {Debenedetti}},\ and\ \bibinfo {author} {\bibfnamefont {F.~H.}\ \bibnamefont
  {Stillinger}},\ }\bibfield  {title} {\bibinfo {title} {Signatures of distinct
  dynamical regimes in the energy landscape of a glass-forming liquid},\
  }\href@noop {} {\bibfield  {journal} {\bibinfo  {journal} {Nature}\ }\textbf
  {\bibinfo {volume} {393}},\ \bibinfo {pages} {554} (\bibinfo {year}
  {1998})}\BibitemShut {NoStop}%
\bibitem [{\citenamefont {Vezirov}\ \emph {et~al.}(2015)\citenamefont
  {Vezirov}, \citenamefont {Gerloff},\ and\ \citenamefont
  {Klapp}}]{vezirov2015manipulating}%
  \BibitemOpen
  \bibfield  {author} {\bibinfo {author} {\bibfnamefont {T.~A.}\ \bibnamefont
  {Vezirov}}, \bibinfo {author} {\bibfnamefont {S.}~\bibnamefont {Gerloff}},\
  and\ \bibinfo {author} {\bibfnamefont {S.~H.}\ \bibnamefont {Klapp}},\
  }\bibfield  {title} {\bibinfo {title} {Manipulating shear-induced
  non-equilibrium transitions in colloidal films by feedback control},\
  }\href@noop {} {\bibfield  {journal} {\bibinfo  {journal} {Soft Matter}\
  }\textbf {\bibinfo {volume} {11}},\ \bibinfo {pages} {406} (\bibinfo {year}
  {2015})}\BibitemShut {NoStop}%
\bibitem [{\citenamefont {Cabriolu}\ \emph {et~al.}(2019)\citenamefont
  {Cabriolu}, \citenamefont {Horbach}, \citenamefont {Chaudhuri},\ and\
  \citenamefont {Martens}}]{cabriolu2019precursors}%
  \BibitemOpen
  \bibfield  {author} {\bibinfo {author} {\bibfnamefont {R.}~\bibnamefont
  {Cabriolu}}, \bibinfo {author} {\bibfnamefont {J.}~\bibnamefont {Horbach}},
  \bibinfo {author} {\bibfnamefont {P.}~\bibnamefont {Chaudhuri}},\ and\
  \bibinfo {author} {\bibfnamefont {K.}~\bibnamefont {Martens}},\ }\bibfield
  {title} {\bibinfo {title} {Precursors of fluidisation in the creep response
  of a soft glass},\ }\href@noop {} {\bibfield  {journal} {\bibinfo  {journal}
  {Soft matter}\ }\textbf {\bibinfo {volume} {15}},\ \bibinfo {pages} {415}
  (\bibinfo {year} {2019})}\BibitemShut {NoStop}%
\bibitem [{\citenamefont {Plimpton}(1993)}]{plimpton1993fast}%
  \BibitemOpen
  \bibfield  {author} {\bibinfo {author} {\bibfnamefont {S.}~\bibnamefont
  {Plimpton}},\ }\href@noop {} {\emph {\bibinfo {title} {Fast parallel
  algorithms for short-range molecular dynamics}}},\ \bibinfo {type} {Tech.
  Rep.}\ (\bibinfo  {institution} {Sandia National Labs., Albuquerque, NM
  (United States)},\ \bibinfo {year} {1993})\BibitemShut {NoStop}%
\bibitem [{\citenamefont {Liu}\ \emph {et~al.}(2018{\natexlab{a}})\citenamefont
  {Liu}, \citenamefont {Ferrero}, \citenamefont {Martens},\ and\ \citenamefont
  {Barrat}}]{liu2018creep}%
  \BibitemOpen
  \bibfield  {author} {\bibinfo {author} {\bibfnamefont {C.}~\bibnamefont
  {Liu}}, \bibinfo {author} {\bibfnamefont {E.~E.}\ \bibnamefont {Ferrero}},
  \bibinfo {author} {\bibfnamefont {K.}~\bibnamefont {Martens}},\ and\ \bibinfo
  {author} {\bibfnamefont {J.-L.}\ \bibnamefont {Barrat}},\ }\bibfield  {title}
  {\bibinfo {title} {Creep dynamics of athermal amorphous materials: a
  mesoscopic approach},\ }\href@noop {} {\bibfield  {journal} {\bibinfo
  {journal} {Soft matter}\ }\textbf {\bibinfo {volume} {14}},\ \bibinfo {pages}
  {8306} (\bibinfo {year} {2018}{\natexlab{a}})}\BibitemShut {NoStop}%
\bibitem [{\citenamefont {Sollich}(2011)}]{sollich2011cecam}%
  \BibitemOpen
  \bibfield  {author} {\bibinfo {author} {\bibfnamefont {P.}~\bibnamefont
  {Sollich}},\ }\bibfield  {title} {\bibinfo {title} {Local strains and yield
  strains in shear flow of amorphous materials, cecam workshop "multiscale
  modelling of amorphous materials: from structure to mechanical properties"}\
  }(\bibinfo {address} {ACAM, Dublin, Ireland},\ \bibinfo {year}
  {2011})\BibitemShut {NoStop}%
\bibitem [{\citenamefont {Patinet}\ \emph {et~al.}(2019)\citenamefont
  {Patinet}, \citenamefont {Barbot}, \citenamefont {Lerbinger}, \citenamefont
  {Vandembroucq},\ and\ \citenamefont {Lema{\^\i}tre}}]{patinet2019origin}%
  \BibitemOpen
  \bibfield  {author} {\bibinfo {author} {\bibfnamefont {S.}~\bibnamefont
  {Patinet}}, \bibinfo {author} {\bibfnamefont {A.}~\bibnamefont {Barbot}},
  \bibinfo {author} {\bibfnamefont {M.}~\bibnamefont {Lerbinger}}, \bibinfo
  {author} {\bibfnamefont {D.}~\bibnamefont {Vandembroucq}},\ and\ \bibinfo
  {author} {\bibfnamefont {A.}~\bibnamefont {Lema{\^\i}tre}},\ }\bibfield
  {title} {\bibinfo {title} {On the origin of the bauschinger effect in
  amorphous solids},\ }\href@noop {} {\bibfield  {journal} {\bibinfo  {journal}
  {arXiv preprint arXiv:1906.09818}\ } (\bibinfo {year} {2019})}\BibitemShut
  {NoStop}%
\bibitem [{\citenamefont {Ozawa}\ \emph {et~al.}(2018)\citenamefont {Ozawa},
  \citenamefont {Berthier}, \citenamefont {Biroli}, \citenamefont {Rosso},\
  and\ \citenamefont {Tarjus}}]{ozawa2018random}%
  \BibitemOpen
  \bibfield  {author} {\bibinfo {author} {\bibfnamefont {M.}~\bibnamefont
  {Ozawa}}, \bibinfo {author} {\bibfnamefont {L.}~\bibnamefont {Berthier}},
  \bibinfo {author} {\bibfnamefont {G.}~\bibnamefont {Biroli}}, \bibinfo
  {author} {\bibfnamefont {A.}~\bibnamefont {Rosso}},\ and\ \bibinfo {author}
  {\bibfnamefont {G.}~\bibnamefont {Tarjus}},\ }\bibfield  {title} {\bibinfo
  {title} {Random critical point separates brittle and ductile yielding
  transitions in amorphous materials},\ }\href@noop {} {\bibfield  {journal}
  {\bibinfo  {journal} {Proceedings of the National Academy of Sciences}\
  }\textbf {\bibinfo {volume} {115}},\ \bibinfo {pages} {6656} (\bibinfo {year}
  {2018})}\BibitemShut {NoStop}%
\bibitem [{\citenamefont {Siebenb{\"u}rger}\ \emph {et~al.}(2012)\citenamefont
  {Siebenb{\"u}rger}, \citenamefont {Ballauff},\ and\ \citenamefont
  {Voigtmann}}]{siebenburger2012creep}%
  \BibitemOpen
  \bibfield  {author} {\bibinfo {author} {\bibfnamefont {M.}~\bibnamefont
  {Siebenb{\"u}rger}}, \bibinfo {author} {\bibfnamefont {M.}~\bibnamefont
  {Ballauff}},\ and\ \bibinfo {author} {\bibfnamefont {T.}~\bibnamefont
  {Voigtmann}},\ }\bibfield  {title} {\bibinfo {title} {Creep in colloidal
  glasses},\ }\href@noop {} {\bibfield  {journal} {\bibinfo  {journal}
  {Physical review letters}\ }\textbf {\bibinfo {volume} {108}},\ \bibinfo
  {pages} {255701} (\bibinfo {year} {2012})}\BibitemShut {NoStop}%
\bibitem [{\citenamefont {Coussot}\ \emph {et~al.}(2006)\citenamefont
  {Coussot}, \citenamefont {Tabuteau}, \citenamefont {Chateau}, \citenamefont
  {Tocquer},\ and\ \citenamefont {Ovarlez}}]{coussot2006aging}%
  \BibitemOpen
  \bibfield  {author} {\bibinfo {author} {\bibfnamefont {P.}~\bibnamefont
  {Coussot}}, \bibinfo {author} {\bibfnamefont {H.}~\bibnamefont {Tabuteau}},
  \bibinfo {author} {\bibfnamefont {X.}~\bibnamefont {Chateau}}, \bibinfo
  {author} {\bibfnamefont {L.}~\bibnamefont {Tocquer}},\ and\ \bibinfo {author}
  {\bibfnamefont {G.}~\bibnamefont {Ovarlez}},\ }\bibfield  {title} {\bibinfo
  {title} {Aging and solid or liquid behavior in pastes},\ }\href@noop {}
  {\bibfield  {journal} {\bibinfo  {journal} {Journal of Rheology}\ }\textbf
  {\bibinfo {volume} {50}},\ \bibinfo {pages} {975} (\bibinfo {year}
  {2006})}\BibitemShut {NoStop}%
\bibitem [{\citenamefont {Liu}\ \emph {et~al.}(2018{\natexlab{b}})\citenamefont
  {Liu}, \citenamefont {Martens},\ and\ \citenamefont {Barrat}}]{liu2018mean}%
  \BibitemOpen
  \bibfield  {author} {\bibinfo {author} {\bibfnamefont {C.}~\bibnamefont
  {Liu}}, \bibinfo {author} {\bibfnamefont {K.}~\bibnamefont {Martens}},\ and\
  \bibinfo {author} {\bibfnamefont {J.-L.}\ \bibnamefont {Barrat}},\ }\bibfield
   {title} {\bibinfo {title} {Mean-field scenario for the athermal creep
  dynamics of yield-stress fluids},\ }\href@noop {} {\bibfield  {journal}
  {\bibinfo  {journal} {Physical Review Letters}\ }\textbf {\bibinfo {volume}
  {120}},\ \bibinfo {pages} {028004} (\bibinfo {year}
  {2018}{\natexlab{b}})}\BibitemShut {NoStop}%
\bibitem [{\citenamefont {Chaudhuri}\ \emph {et~al.}(2012)\citenamefont
  {Chaudhuri}, \citenamefont {Berthier},\ and\ \citenamefont
  {Bocquet}}]{chaudhuri2012inhomogeneous}%
  \BibitemOpen
  \bibfield  {author} {\bibinfo {author} {\bibfnamefont {P.}~\bibnamefont
  {Chaudhuri}}, \bibinfo {author} {\bibfnamefont {L.}~\bibnamefont
  {Berthier}},\ and\ \bibinfo {author} {\bibfnamefont {L.}~\bibnamefont
  {Bocquet}},\ }\bibfield  {title} {\bibinfo {title} {Inhomogeneous shear flows
  in soft jammed materials with tunable attractive forces},\ }\href@noop {}
  {\bibfield  {journal} {\bibinfo  {journal} {Physical Review E}\ }\textbf
  {\bibinfo {volume} {85}},\ \bibinfo {pages} {021503} (\bibinfo {year}
  {2012})}\BibitemShut {NoStop}%
\end{thebibliography}%

\end{document}